\documentstyle[psfig]{mn}
\def\be{\begin{equation}}
\def\ee{\end{equation}}
\def\bea{\begin{eqnarray}}
\def\eea{\end{eqnarray}}

\def\bfig{\begin{figure}}
\def\efig{\end{figure}}
\def\btab{\begin{table}}
\def\etab{\end{table}}

\def\eg{{\rm e.g.\ }}
\def\etc{{\rm etc.\ }}
\def\ie{{\rm i.e.\ }}
\def\Fig{Figure}

\def\Eq{equation}
\def\Eqs{\Eq{}s}

\def\Sec{Section}

\def\dd{\hbox{\rm d}}
\def\spose#1{\hbox to 0pt{#1\hss}}
\def\approxlt{\mathrel{\spose{\lower 3pt\hbox{$\sim$}}
	\raise 2.0pt\hbox{$<$}}}
\def\approxgt{\mathrel{\spose{\lower 3pt\hbox{$\sim$}}
	\raise 2.0pt\hbox{$>$}}}
\def\approxpropto{\mathrel{\spose{\lower 3pt\hbox{$\sim$}}
	\raise 2.0pt\hbox{$\propto$}}}

\def\refindent{\par \noindent \hang}
\def\paper#1#2#3#4#5{\refindent #1, #2, #3, #4, #5}

\def\confoe#1#2#3#4#5#6{\refindent #1, #2, in #4, ed, #3. #5, p.~#6}

\def\and{, }
\def\MN{MNRAS}

\def\bfit{\beta_{\rm fit}}
\def\icm{ICM}
\def\rvec{{\bf r}}
\def\uvec{{\bf u}}

\begin{document}

\title[Mass constraints from cooling flows]
{Mass constraints from multiphase cooling flow models}
\author[P. A. Thomas]
{Peter A. Thomas \\
Astronomy Centre, School of Chemistry, Physics and Environmental Science, 
University of Sussex, Falmer, Brighton BN1 9QH}

\date{Accepted ---. Received ---; in original form ---}

\maketitle

\begin{abstract}
I review the multiphase cooling flow equations that reduce to a
relatively simple form for a wide class of self-similar density
distributions described by a single parameter, $k$.  It is shown that
steady-state cooling flows are \emph{not} consistent with all possible
emissivity profiles which can therefore be used as a test of the
theory.  In combination, they provide strong constraints on the
temperature profile and mass distribution within the cooling radius.
The model is applied to ROSAT HRI data for 3 Abell clusters.  At one
extreme ($k\sim1$) these show evidence for cores in the mass
distribution of size 110--140$\,h_{50}^{-1}$kpc and have temperatures
which decline towards the flow centre.  At the other
($k\mapsto\infty$), the mass density and gas temperature both rise
sharply towards the flow centre.  The former are more consistent with
measured temperatures, which suggests that the density distribution in
the intracluster medium contains the minimum possible mixture of
low-density components.
\end{abstract}


\section{Introduction}

The multiphase nature of the intracluster medium (\icm) in cooling
flows was demonstrated a decade ago when deprojections of X-ray
surface-brightness profiles showed that mass cools and is deposited
from the flow in a distributed manner, $\dot{M}\approxpropto r$ (\eg
Thomas, Fabian \& Nulsen 1987).  However, the complexity of the theory
and lack of data with high spatial resolution means that the
single-phase approximation is still widely adopted.  In this paper I
show how the use of self-similar density distributions can lead to
a relatively simple form for the multiphase cooling flow equations
whilst spanning the whole range of expected behaviours in the more
general case.  In combination with the emissivity profile (for a
spherically-symmetric flow), the equations can be solved to yield the
gas temperature and mass profiles within the cooling flow.

In Section~\ref{sec:dd}, I introduce the concept of the volume
fraction to describe the distribution of density phases in the \icm.
The self-similar forms of the volume fraction are determined and shown to
span all reasonable behaviours for the more general situation.  The
steady-state, self-similar form of the cooling flow equations are
derived in \Sec~\ref{sec:theory} and it is shown how these can be
solved in the spherically-symmetric case if the emissivity profile is
known.  The behaviour of the solutions is examined in detail: not
every emissivity profile is consistent with these models.  In
\Sec~\ref{sec:abell}, the theory is applied to ROSAT HRI data for
three Abell clusters with some success.  Finally, the results are
summarised and further discussed in \Sec~\ref{sec:conc}.

\section{The form of the density distribution}
\label{sec:dd}

The theory of multiphase cooling flows was set out by Nulsen (1986).
He introduced the concept of the volume fraction,
$f(\rho,\rvec,t)$, such that $f\,\dd\rho$ is the fractional volume
occupied by phases with densities in the range $\rho$ to
$\rho+\dd\rho$.  It is assumed that the phases comove but are
thermally isolated from one another---these can be regarded as
empirical facts as otherwise the flow would rapidly evolve to a
single-phase state.

Given these assumptions then it is relatively straightforward to
derive an equation for the evolution of the volume fraction (see
Appendix).  Writing
\be
f={(2-\alpha)\over\rho_0}w^{(4-\alpha)/(2-\alpha)}g(w,\rvec,t),
\ee
where
\be
w=\left(\rho_0\over\rho\right)^{2-\alpha}
\ee
and $\rho=\rho(\rvec,t)$, then
\be
{\dot{g}\over g}+(3-\alpha){\dot{\rho_0}\over\rho_0}-
{2-\alpha\over\gamma}{\dot{P}\over P}+\nabla.\uvec=0
\label{eq:gdot}
\ee
where the dot refers to a covariant derivate, following the flow.

In general $g$ is a complicated function of position and time.
However, we can look for solutions in which $g$ has a constant
functional form, $g=g(w)$.  Only the first term in \Eq~\ref{eq:gdot}
depends upon $w$.  Hence we require that $\dot{g}\equiv\dot{w}\,\dd
g/\dd w\propto g$.  There are two kinds of solution:
\begin{enumerate}
\item $g_\infty\propto\exp(-w)$.  This
is the most extended distribution which is convectively stable (it
gives $P\propto\rho_0^\gamma$).  It includes phases of arbitrarily low
density.
\item $g_k\propto(1-w)^{k-1}$, $0<w<1$; $k\geq1$.  
These solutions possess a minimum density, $\rho\geq\rho_0$.  $k=1$ is
the least extended, consisting solely of the power-law cooling tail.
As $k\mapsto\infty$ the solutions resemble $g_\infty$.
\end{enumerate}

For other forms of $g$ one must resort to numerical integration to
follow their evolution.  Thomas (1988b) looked at the steady-state
evolution of a range of distributions with a sharp cut-off at low
densities and reached the following conclusions:
\begin{itemize}
\item All distributions develop a high-density tail,
$f\sim\rho^{-(4-\alpha)}$, as they cool.
\item Sufficiently narrow distributions resemble the pure power-law
$g_1$ by the time they begin to be deposited.
\end{itemize}

$g_1$ and $g_\infty$ bound all reasonable solutions of the cooling
flow equations, be they self-similar in form or not.  In principle,
$k$ could be less than unity (as $k\mapsto0$ the flow tends to the
homogeneous case) but it is difficult to see how such distributions
could arise.  If the large density variations inferred within the
cooling radius result from amplification of an initially much narrower
distribution, then values of $k$ of order unity are to be preferred.
Without a plausible formation history for the ICM, however, it is
better to leave $k$ as a free parameter to be determined empirically.
I will argue below that low values of $k$ fit the observations of
cooling flows in Abell clusters better than high values.

\section{Reconstruction of cluster mass profiles}
\label{sec:theory}

The self-similar density distributions derived in \Sec~\ref{sec:dd}
lead to particularly simple forms of the steady-state cooling flow
equations.  I derive these below and then show in the next section how
they can be combined with the emissivity profile to provide strong
constraints of the mass distribution within the cooling radius.

\subsection{The equations}

Substituting the functional form of $g_k$ into 
\Eqs~\ref{eq:mass1}, \ref{eq:beta} and \ref{eq:gdot} we see that the
self-similar forms of the cooling flow equations are:
\be
{\dot{\rho_0}\over\rho_0}-{1\over\gamma}{\dot{P}\over P}-
{\beta\over(2-\alpha)k}=0
\ee
and
\be
{\dot{\rho_0}\over\rho_0}+\nabla.\uvec+\beta=0,
\ee
where
\be
\beta=(2-\alpha)k{\gamma-1\over\gamma}{n_0^2\Lambda(T_0)\over P}.
\ee
To these may be added the equation of hydrostatic support,
\be
\nabla\Phi+{\nabla P\over\bar\rho}=0,
\ee
where $\Phi$ is the gravitational potential.  I assume here that the inflow
is subsonic---this turns out to be a good approximation in all
multiphase cooling flow models.

In a steady-state and spherical symmetry the above equations reduce to
\be
{\dd\ln P\over\dd\ln r}=-2\Sigma,
\ee
\be
{\dd\ln\bar{\rho}\over\dd\ln r}=-{2\over\gamma}\Sigma-{\tau\over2-\alpha},
\ee
and
\be
{\dd\ln u\over\dd\ln r}=-2+{2\over\gamma}\Sigma+
\left({1\over2-\alpha}+k\right)\tau,
\ee
where
\be
\Sigma={GM\over2r}.{\mu m_H\over k_BT}
\ee
is the ratio of the virial to the thermal temperatures, and
\be
\tau=(2-\alpha){\gamma-1\over\gamma}{n_0^2\Lambda(T_0)\over P}
{r\over u}={1\over k}{\dd\ln\dot{M}\over\dd\ln r}
\ee
is the ratio of the inflow time to the constant-pressure cooling time.

Although there appear to be three equations here, the dimensionless
ratios $\Sigma$ and $\tau$ are the only important variables.  The
third equation merely acts as a scaling (for fixed $\tau$,
$\bar{\rho}^{2-\alpha}\propto p^{1-\alpha}u$).  Hence the physics can
be captured in just two equations:
\be
{\dd\ln\Sigma\over\dd\ln r}=\chi-1+2{\gamma-1\over\gamma}\Sigma-
{\tau\over2-\alpha}
\ee
and
\be
{\dd\ln\tau\over\dd\ln r}=3-
{2\over\gamma}\big[(3-\alpha)-\gamma(1-\alpha)\big]\Sigma-
\left({3-\alpha\over2-\alpha}+k\right)\tau
\ee
where $\chi\equiv\dd\ln M/\dd\ln r$.  The $g_\infty$
equations can be recovered by letting $k\mapsto\infty$ and using
$k\tau$ in place of $\tau$ as the second dimensionless variable.

The usual way of proceeding when solving the single-phase equations is
to pick a functional form for the mass profile, $\chi$, and mass
deposition rate, $\beta$, then to solve for $\Sigma$ and $\tau$.  From
this one can generate an emissivity profile, $\xi(r)$, for comparison
with the data.  Here I adopt a different approach: because $\beta$ is
fixed (for a particular value of $k$) in the multiphase models, one
can specify $\xi(r)$ and \emph{determine} the mass profile.

Suppose that $\bfit\equiv-(1/6)\dd\ln\xi/\dd\ln r$ is known.  Then
\begin{eqnarray}
{\dd\ln\tau\over\dd\ln r} = 3 &-& 
6{(3-\alpha)-\gamma(1-\alpha)\over2-\alpha+\alpha\gamma}\bfit \nonumber \\
{}&-&\left({2\gamma\over(2-\alpha)(2-\alpha+\alpha\gamma)}+k\right)\tau.
\label{eq:tau}
\end{eqnarray}
Furthermore this is an eigenvalue problem: requiring that the solution
extend to $r=0$ fixes the outer boundary condition.  Hence we can
solve for $\tau$, $\Sigma$ and $\chi$.

\subsection{The behaviour of solutions}
\label{sec:behaviour}

We can get a good idea of the behaviour of the solutions to
\Eq~\ref{eq:tau} by looking at the case of constant $\bfit$.  Imposing the
physical constraints $\Sigma\geq 0$ (\ie a non-negative temperature)
and $\chi\geq0$ (\ie mass constant or increasing with radius)
restricts $\bfit$ to lie in the range
\be
{3\over2(5+3\,k)}\leq\bfit\leq{80+21\,k\over120+36\,k}
\label{eq:krange}
\ee
(in this expression and henceforth I set $\gamma=5/3$ and $\alpha=0.5$
rather than including them explicitly).  Thus steep emissivity
profiles, $\bfit\approxgt0.65$, are incompatible will all steady-state
cooling flow models (larger values can occur, however, outside the
cooling radius).  In addition, the inner value of $\bfit$ can be used
to constrain $k$: flat cores are inconsistent with small values of $k$
and a central decrease in emissivity would be incompatible with all
models.  If we assume that the virial temperature drops
(\ie $\Sigma\mapsto0$) within the cluster core, then the central value
of $\bfit$ provides a \emph{measure} of $k$.

\begin{figure*}
\parbox{8.7truecm}{
\psfig{figure=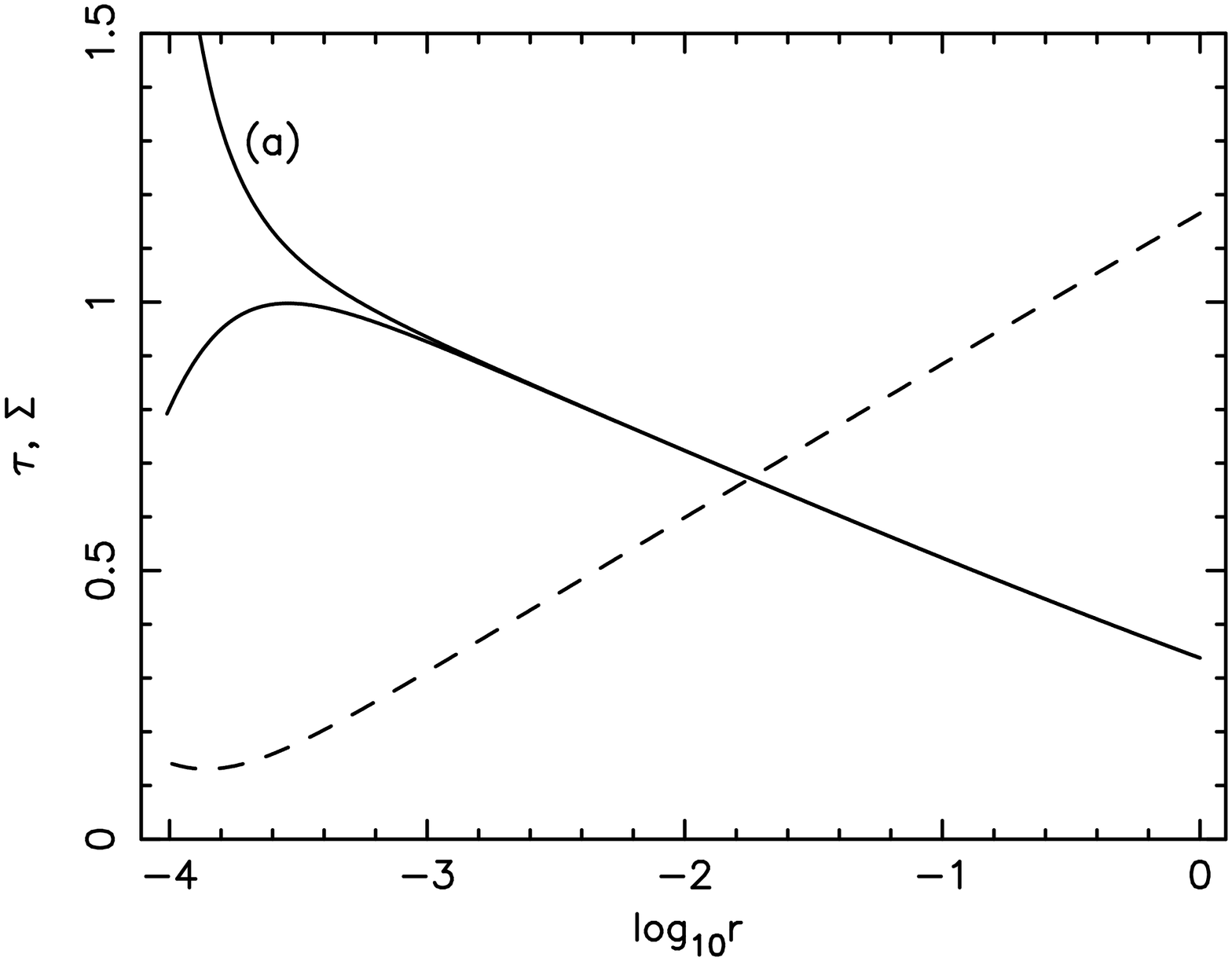,width=8.7cm,angle=270}
}\parbox{8.7truecm}{
\psfig{figure=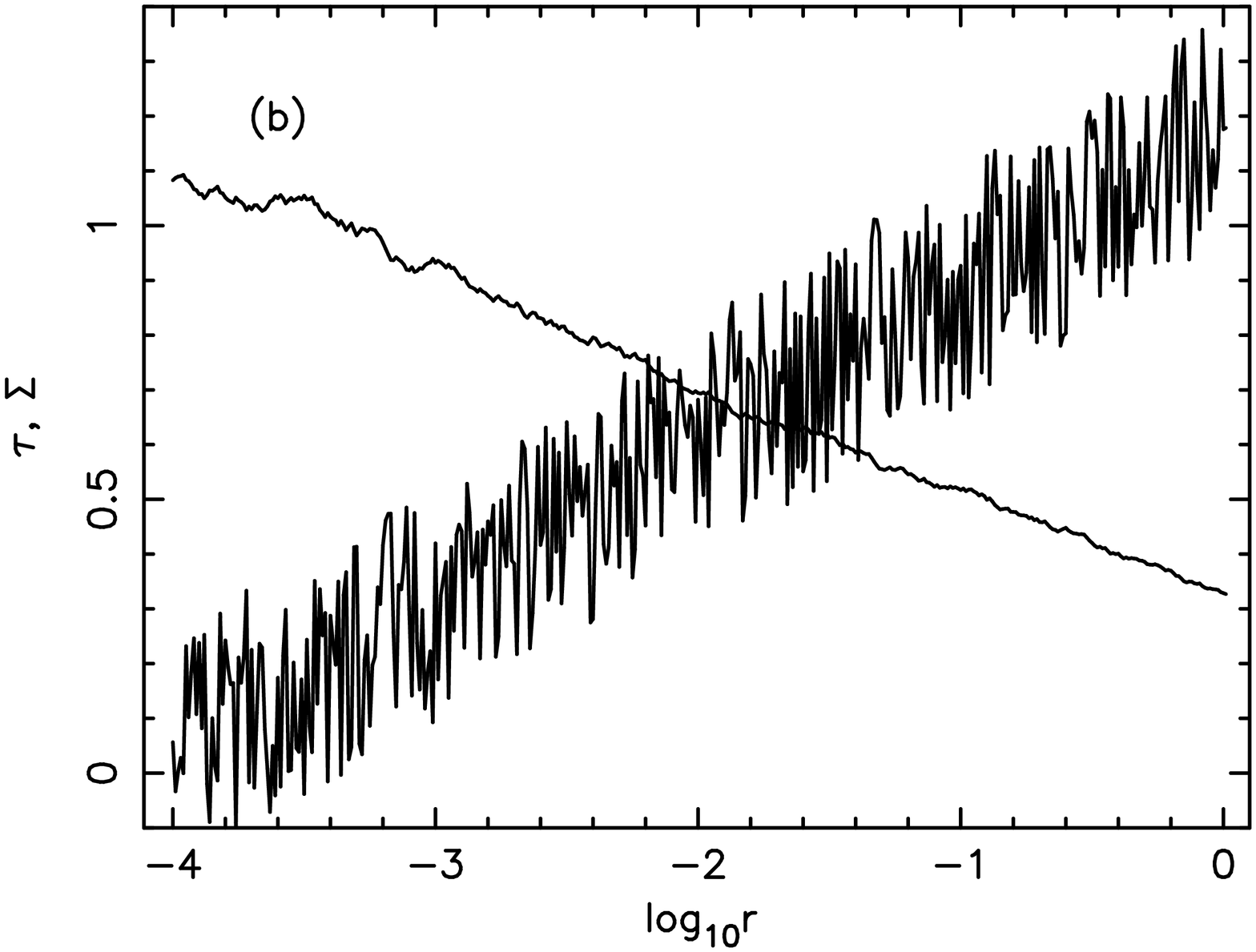,width=8.7cm,angle=270}
}
\caption{Solutions to the multiphase cooling flow equation,
\Eq~\ref{eq:tau}, for simple test profiles: (a)
$\bfit=0.6+0.1\,\log_{10}r$, (b) $\bfit=0.6+0.1\,\log_{10}r\pm0.1$.
The curves which decline and increase with radius represent solutions
for $\tau$ and $\Sigma$, respectively.}
\label{fig:test}
\end{figure*}

The numerical solutions of \Eq~\ref{eq:tau} in the case of
non-constant $\bfit$ are illustrated in \Fig~\ref{fig:test}.  Here I
have set $\bfit=0.6+0.1\,\log_{10}r$ and $k=1$.  Panel
(a) shows the speed at which solutions diverge away from the desired
one (\ie the one which extends to $r=0$) as $r$ decreases.  This
ensures that the solutions are insensitive to the inner boundary
condition and are stable to small variations in the value of $\bfit$
in the inner bin (which is most likely to be affected by the
point-spread function, uncertain absorption correction, \etc).  The
difference in the two values of $\tau$ at $r=10^{-3}$ in the figure is
less than one percent.  Panel (b) shows that the solutions are able to
cope with quite large variations in $\bfit$: in this case 100
fluctuations drawn from a uniform distribution of extent $\pm0.1$ have
been added to each decade in $r$.  Despite the fact that $\bfit$ now
varies outside both ends of the physical range described in
\Eq~\ref{eq:krange}, it remains true to the smooth solution.
In this paper we will fit smooth functions to observed emissivity
profiles before attempting to solve the cooling flow equations.

\section{Application to Abell clusters}
\label{sec:abell}

I will now apply the above theory to three cooling flow clusters:
A\,85, A\,745 and A\,2025.  The ROSAT HRI data were kindly supplied to
me by Clovis Peres in the form of deprojected density and temperature
profiles.  This has the advantage that conversion of the counts into
emissivity, including correction for the spectral response and
absorption, has already been included.  Provided that the cooling flow
solutions resemble the deprojected one, then these corrections will
hold good.

I will consider each of the clusters in turn.

\subsection{A\,85}
\label{sec:a85}

\begin{figure}
\psfig{figure=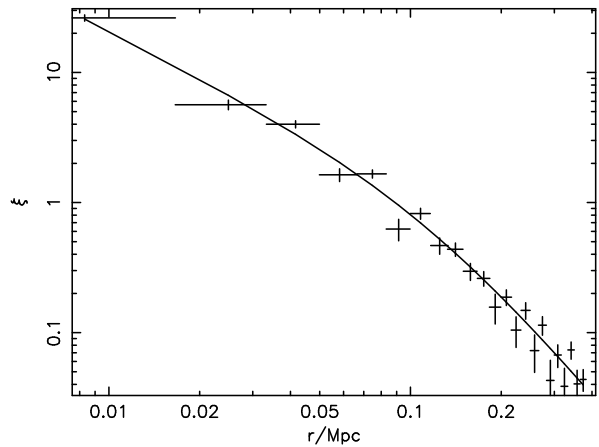,width=8.7cm,angle=270}
\caption{The emissivity profile for A85 in 12 arcsec bins.  The solid
line shows a broken power-law fit as described in the text.}
\label{fig:a85xi}
\end{figure}

\begin{figure*}
\parbox{8.7truecm}{
\psfig{figure=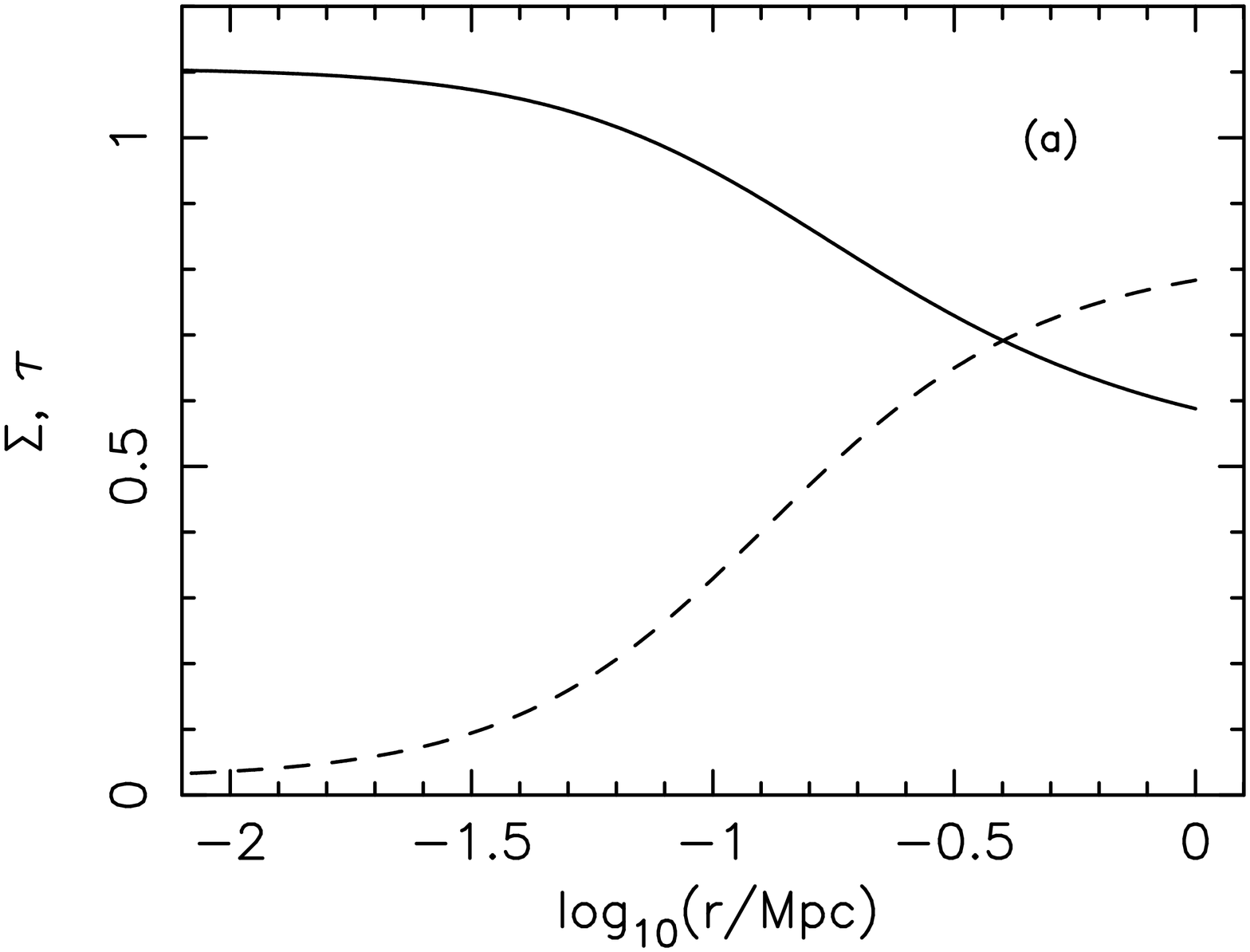,width=8.7cm,angle=270}
}\parbox{8.7truecm}{
\psfig{figure=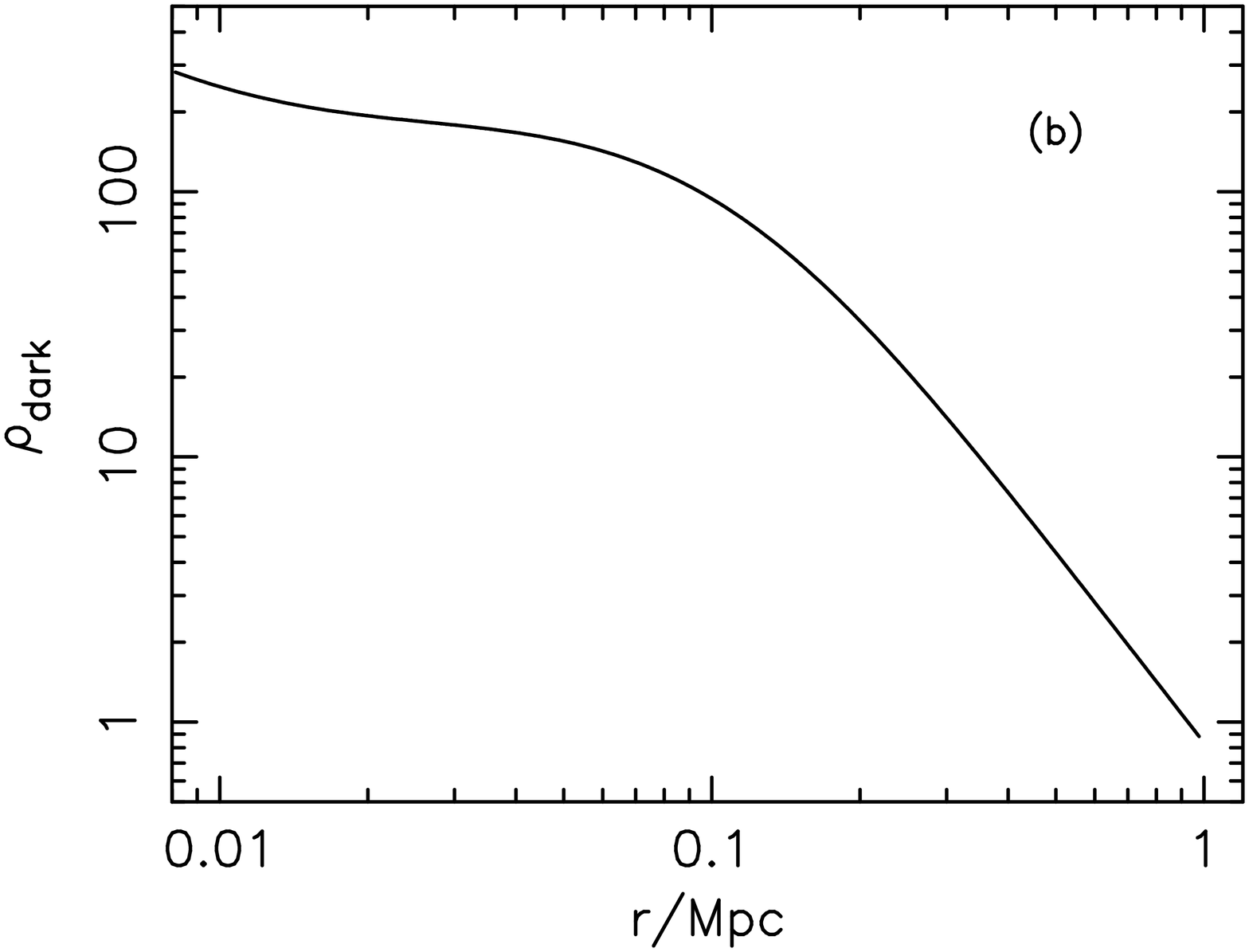,width=8.7cm,angle=270}
}
\caption{The $k=1$ cooling flow solution for A85; (a) $\Sigma$ (dashed
line) and $\tau$ (solid line), (b) the density of the gravitating matter.}
\label{fig:a85k001}
\end{figure*}

The emissivity profile is shown in \Fig~\ref{fig:a85xi}.  There are 23
annular bins, each 12 arcsec wide, of which the inner 9
have cooling times less than $1.33\times10^{10}$yr (throughout this
paper I take $H_0=50\,$km\,s$^{-1}$Mpc$^{-1}$).
The profile is well-fit by a simple broken power-law fit,
\be
\xi\propto\left[\left(r\over0.12\,{\rm Mpc}\right)^{1.18}+
\left(r\over0.12\,{\rm Mpc}\right)^{2.83}\right]^{-1}.
\ee
The asymptotic slope of $\xi$ as $k\mapsto0$ is very close to the minimum
permitted for $k=1$ which suggests that the solution will require an
inner core in the mass distribution.  This is illustrated in
\Fig~\ref{fig:a85k001}.  Note that $\Sigma$
drops very close to zero at $r=10\,$kpc (it tends to a small constant
value within this radius).  The gravitational density profile (\ie
that of the total mass, not just the gas) is well-fit at radii greater
than 20\,kpc by a King model,
\be
\rho_{\rm grav}
\propto\left[1+\left(r\over0.11\,\mbox{Mpc}\right)^2\right]^{-1.25}.
\ee
Within this radius the density is poorly constrained.  Although it
appears to rise, only a small change in the slope of $\xi$
would cause it to level off or even fall---all we know for sure is
that the virial temperature becomes very small.  Note also that there
is only one bin within 20\,kpc and this one is most likely to be
affected by smoothing by the point-spread function, uncertain correction
for excess absorption, etc.

The temperature of the gas is approximately constant outside the core
radius, but drops by a factor of five in to 10\,kpc.  A temperature
decline in the centre of clusters is typical of cooling flows observed
by ASCA.

Note that the slope of the mass-deposition profile, $\tau$, is close
to unity within the cooling radius, $r_{\rm cool}\approx150\,$kpc.
This radius is not a special one for our solutions as we have assumed
the cooling flow solution holds everywhere.  For this reason the
asymptotic slope of the gravitational density profile at larger radii
should be taken with a pinch of salt.

\begin{figure*}
\parbox{8.7truecm}{
\psfig{figure=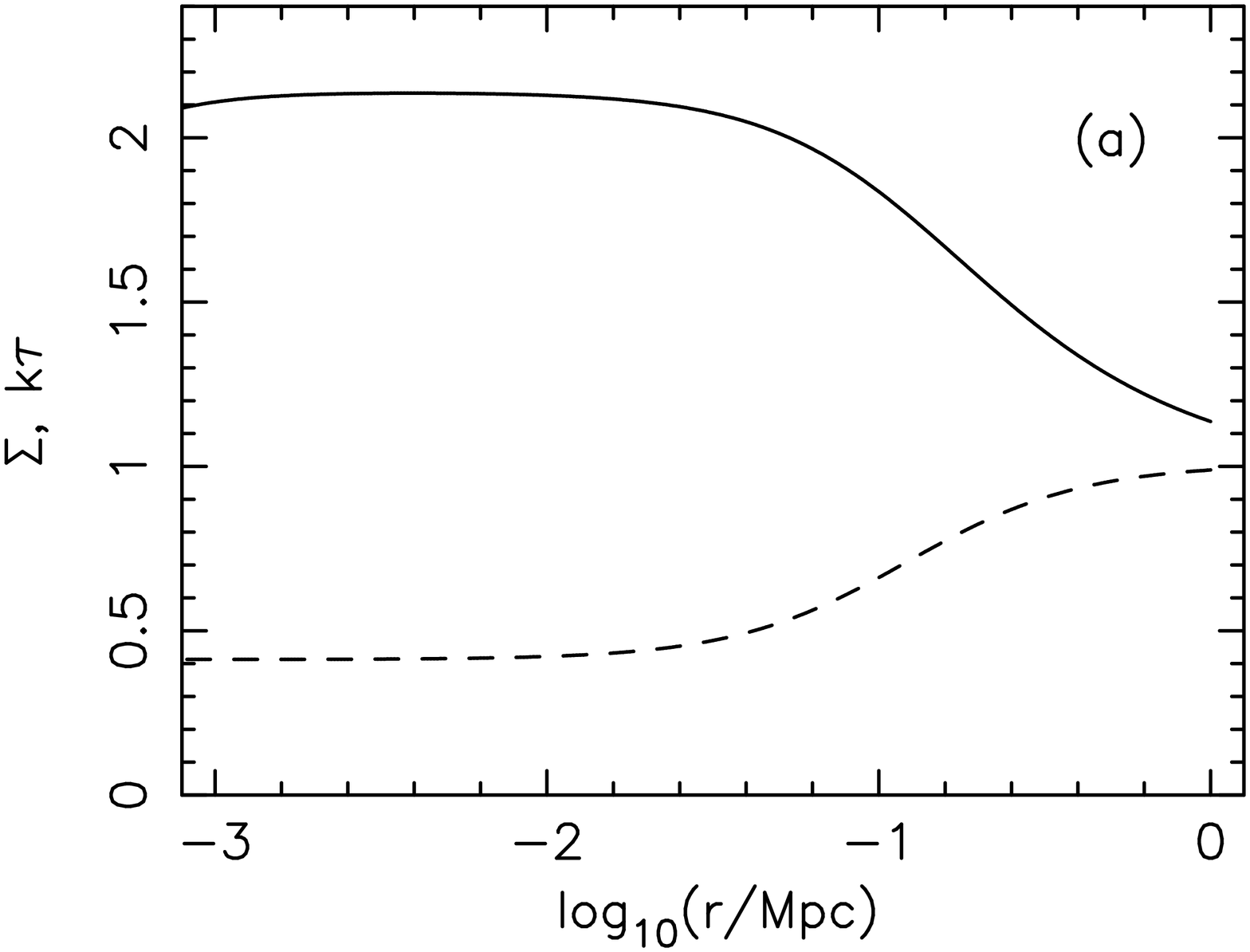,width=8.7cm,angle=270}
}\parbox{8.7truecm}{
\psfig{figure=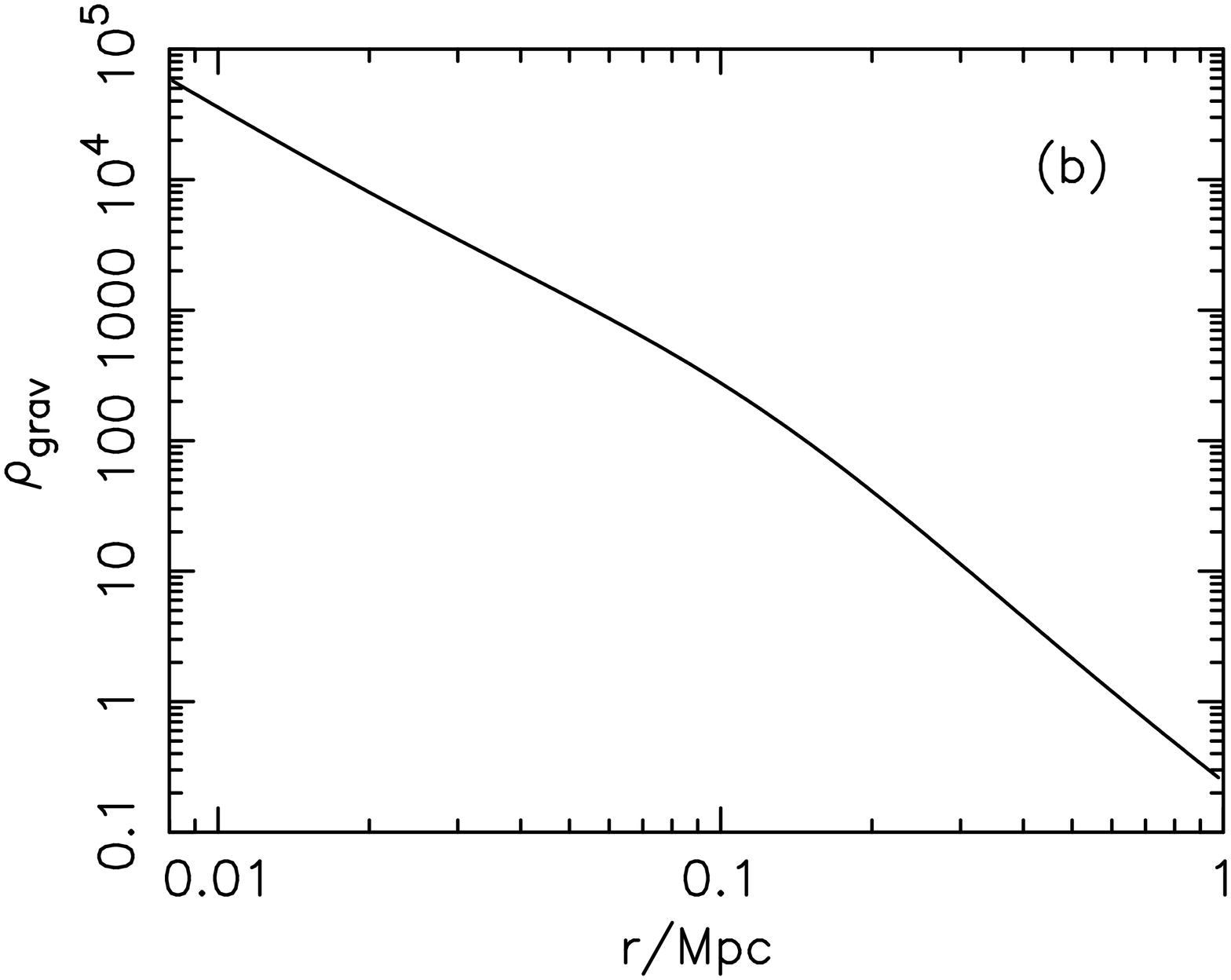,width=8.7cm,angle=270}
}
\caption{The $k=\infty$ cooling flow solution for A85; (a) $\Sigma$ (dashed
line) and $\tau$ (solid line), (b) the density of the gravitating matter.}
\label{fig:a85k100}
\end{figure*}

The corresponding solution for $k=\infty$ is shown in \Fig~\ref{fig:a85k100}.
There in no core in the gravitational mass profile in this
case, with $\rho_{\rm dark}$ rising as $r^{-2}$ all the way into
10\,kpc.  This is reflected in the temperature profile, however, which
also rises by a factor of 5 between 150 and 10\,kpc, unlike any
observed cluster.

\subsection{A\,745}
\label{sec:a745}

\begin{figure}
\psfig{figure=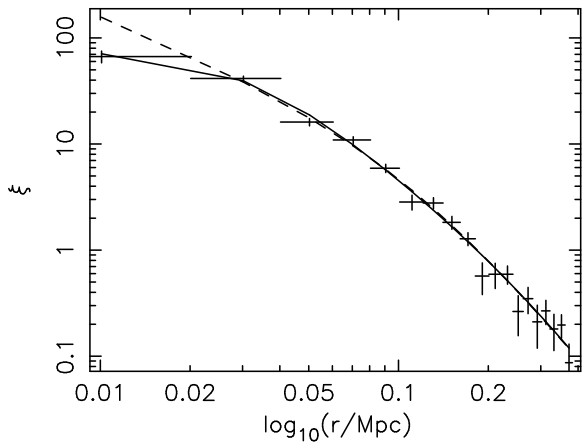,width=8.7cm,angle=270}
\caption{The emissivity profile for A745 in 8 arcsec bins.  The solid
line shows an analytic fit to the data and the dashed line
shows the emissivity profile corresponding to a King-law
density profile, as described in the text.}
\label{fig:a745xi}
\end{figure}

The emissivity profile, shown in \Fig~\ref{fig:a745xi}, has 19 bins of
width 8 arcsec, with 9--12 bins interior to the cooling radius
($r_{\rm cool}\approx180$--240\,kpc).  The first thing to note is that
$\xi$ flattens considerably in the innermost bin.  This is
incompatible with density distributions with small values of $k$.  In
fact the fit shown by the solid line,
\be
\xi\propto\left[1+\left(r\over31\,{\rm kpc}\right)^{2.28}+
\left(r\over89\,{\rm kpc}\right)^{4.09}\right]^{-1}
\ee
admits solutions which extend to $r=0$ only for $k=\infty$.
Unfortunately, just as for A\,85, large values of $k$ give steeply
rising dark-matter density and temperature profiles at small radii.

If we ignore the inner bin, however, then it is possible to find
solutions for all values of $k$.  The dotted line in
\Fig~\ref{fig:a745xi} shows the emissivity profile which corresponds
to a mass density
\be
\rho_{\rm grav}
\propto\left[1+\left(r\over0.11\,\mbox{Mpc}\right)^2\right]^{-1.57}.
\ee
and $k=1$.

\subsection{A\,2029}
\label{sec:a2029}

\begin{figure}
\psfig{figure=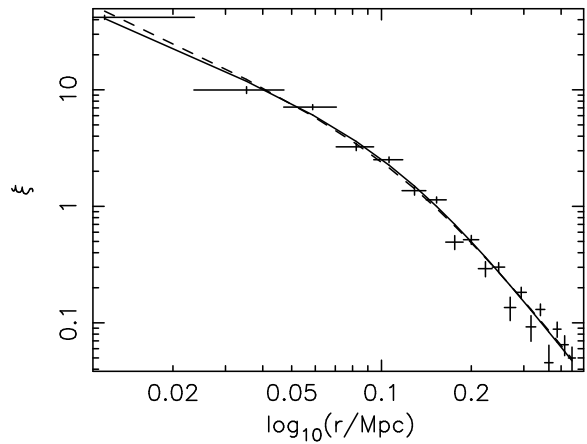,width=8.7cm,angle=270}
\caption{The emissivity profile for A2029 in 12 arcsec bins.  The solid
line shows an analytic fit to the data and the dashed line
shows the emissivity profile corresponding to an analytic
density profile, as described in the text.}
\label{fig:a2029xi}
\end{figure}

\begin{figure*}
\parbox{8.7truecm}{
\psfig{figure=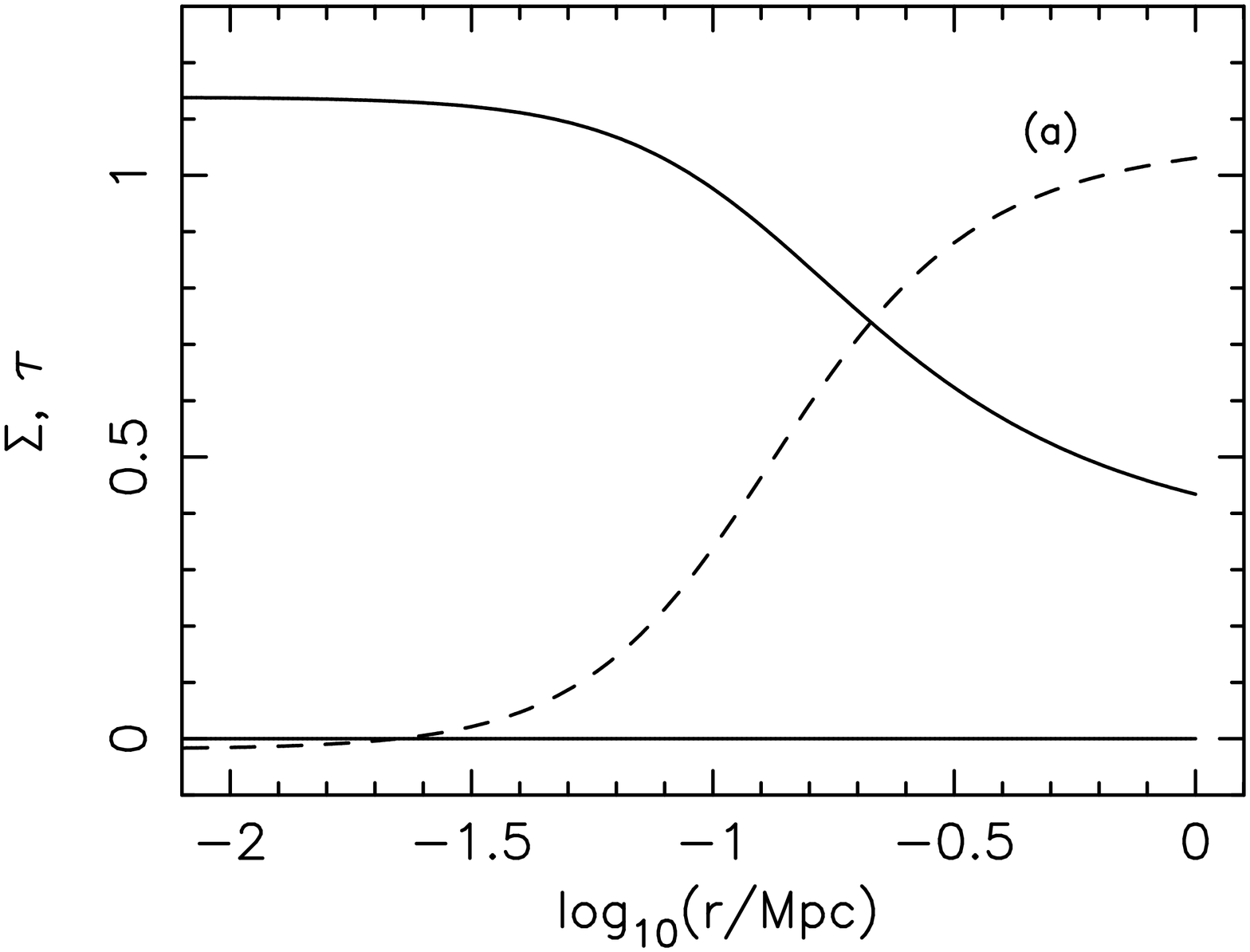,width=8.7cm,angle=270}
}\parbox{8.7truecm}{
\psfig{figure=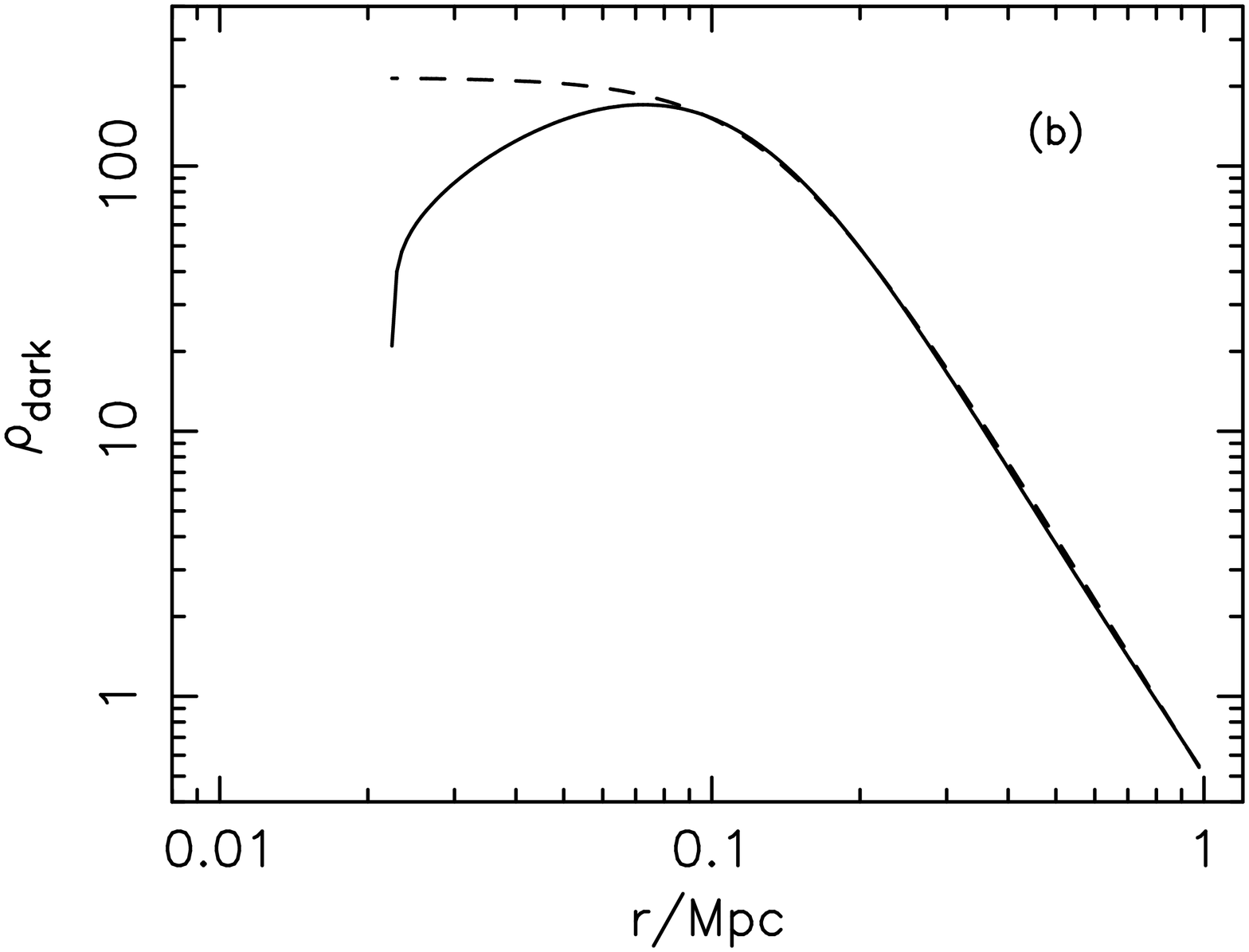,width=8.7cm,angle=270}
}
\caption{The $k=1$ cooling flow solution for A2029; (a) $\Sigma$ (dashed
line) and $\tau$ (solid line), (b) the density of the gravitating
matter (solid line) and an analytic fit as described in the text
(dashed line).}
\label{fig:a2029k001}
\end{figure*}

For A\,2029 there are 7--9 bins of width 12 arcsec within the cooling
radius, $r_{\rm cool}approx170$--210\,kpc.  The emissivity profile
\be
\xi\propto\left[\left(r\over0.13\,{\rm Mpc}\right)^{1.09}+
\left(r\over0.13\,{\rm Mpc}\right)^{3.34}\right]^{-1}.
\ee
shown in \Fig~\ref{fig:a2029xi} is again too shallow in the central
bin, but the inconsistency is this time so slight that it strengthens
the case for $k=1$ (as discussed in \Sec~\ref{sec:behaviour} the slope
of the emissivity profile within the cluster core provides a measure
of $k$).
\Fig~\ref{fig:a2029k001}a shows that
$\Sigma$ drops to a value close to zero within 30\,kpc.  This
indicates that the virial temperature has sunk well below the gas
temperature (\ie the flow is isobaric within this radius).  
This is
reflected in \Fig~\ref{fig:a2029k001}b which shows the corresponding
density profile.  The latter is well-fit at radii greater than 80\,kpc
by a mass-density
\be
\rho_{\rm grav}
\propto\left[1+\left(r\over0.14\,\mbox{Mpc}\right)^3\right]^{-1}
\ee
but declines rapidly within this radius.  Only a
small change in the emissivity profile is required to generate
solutions in which the gravitational mass has a constant core density,
as indicated by the dotted line in \Fig~\ref{fig:a2029xi} which
reproduces the above density profile exactly.

\section{Conclusions and discussion}
\label{sec:conc}

\begin{itemize}
\item I have rederived the cooling flow equations for self-similar
density distributions.  Two of these in particular are expected to
bound the behaviour of all possible flows.
\item The steady-state cooling flow equations are \emph{not}
compatible will all conceivable emissivity profiles.  Thus they can be
used as a test of the theory.
\item The solutions provide bounds on $M(r)$ within the cooling radius
and given $k$ can be used to measure $M(r)$.
\item ROSAT HRI surface brightness data from three Abell clusters are
best-fit by models in which $k\approx1$.  The measured core radii are
then 110--140\,$h_{\rm 50}^{-1}\,$kpc.  Larger values of $k$ give
smaller (or non-existent) core radii, but have gas temperatures which
rise sharply at small radii, in contrast with the observations.
\end{itemize}

It is unfortunate that the core radii deduced for the matter
distributions for the Abell clusters under investigation are only
slightly smaller than the radii of the cooling flows.  It is possible
that the self-similar, steady-state assumption is a poor one at the
edge of the flow (it is also possible that it is valid well beyond the
cooling radius) and that models could be found which were consistent
with a much larger core radius.

Ideally, one would like a much better resolution at smaller radii, so
that the tendency of the emissivity profile to a constant slope,
indicative of the particular value of $k$, could be checked.  This
should be possible for some of the closer clusters, such as Virgo.

Note that the solutions for $\Sigma$ and $\tau$ given in this paper do
not depend upon the normalisations of the gas and total gravitational
masses.  If desired, these can be determined by fixing the overall
temperature and luminosity.  The analysis of Gunn \& Thomas (1996)
shows that the gas density will be slightly lower and the total mass
density slightly higher than in the equivalent single-phase analysis.

Cooling flows in individual galaxies are much better resolved than
those in clusters and so look like promising candidates for the kind
of modelling discussed here.  However, the lower temperatures leads to
complications such as a varying slope, $\alpha$, for the cooling
function, and larger corrections for absorption and emission lying
outside the pass-band of the detector.  There may also be sources of
mass and energy injection into the flow.

\section*{Acknowledgments}

This paper was prepared using the facilities of the STARLINK minor
node at Sussex.  It was written while PAT was holding a Nuffield
Foundation Science Research Fellowship.

\section*{References}
\paper{Gunn K. F., Thomas P. A.}{1996}{\MN}{281}{1133}
\paper{Nulsen P. E. J.}{1986}{\MN}{221}{377}
\confoe{Thomas P. A.}{1988a}{NATO ASI Cooling flows in
   clusters and galaxies}{Fabian A. C.}{Kluwer, Dordrecht}{361}
\paper{Thomas P. A.}{1988b}{\MN}{235}{315}
\paper{Thomas P. A., Fabian A. C., Nulsen P. E. J. N.}{1987}{\MN}{228}{973}

\appendix

\section{Derivation of the multiphase cooling flow equations}

We assume an emulsion of density phases which comove with the flow.
The distribution is described by the volume fraction,
$f(\rho,\rvec,t)$, such that $f\,\dd\rho$ is the fractional volume
occupied by phases with densities in the range $\rho$ to
$\rho+\dd\rho$. Then $\int f\,\dd\rho=1$, and the mean density is
$\bar{\rho}=\int f\rho\,\dd\rho$.

Mass conservation gives
\be
{\partial\over\partial t}(\rho f)+\nabla.(\uvec\rho f)+
{\partial\over\partial\rho}(\dot\rho\rho f)=0,
\label{eq:mass}
\ee
where $\uvec$ is the rate of change of position and $\dot\rho$ is the
rate of change of density following the flow.  The final term in
\Eq~\ref{eq:mass} is the equivalent in density space of the
divergence in velocity space.

Integrating over all densities we obtain
\be
{\dot{\bar{\rho}}\over\bar{\rho}}+\nabla.\uvec+\beta=0,
\label{eq:mass1}
\ee
where
\be
\beta\equiv{1\over\bar{\rho}}\lim_{\rho\mapsto\infty}(\dot{\rho}\rho f).
\label{eq:beta}
\ee
This is equivalent to the usual single-phase equation (\eg Thomas
1988a) except that the mass deposition rate is specified in terms of
$f$ rather than being a free parameter.

To find how the volume fraction changes with time we use the energy
equation,
\be
\rho\dot{s}=-n^2\Lambda,
\label{eq:sdot}
\ee
where $s$ is the entropy and $n^2\Lambda$ is the radiated power per unit
volume.  For a fully-ionised plasma
\be
s\equiv{1\over\gamma-1}{k_B\over\mu m_H}\ln\left(P\over\rho^\gamma\right).
\label{eq:entropy}
\ee
where $P$ is the pressure, $k_B$ is the Boltzmann constant, $\mu m_H$
is the mass per particle and $\gamma=5/3$.  Then
\be
{\dot{\rho}\over\rho}={1\over\gamma}{\dot{P}\over P}+
{\gamma-1\over\gamma}{n^2\Lambda\over P}.
\label{eq:energy}
\ee

Over a wide temperature range appropriate to clusters the cooling
function can be approximated by a power-law, $\Lambda\propto
T^\alpha$, where $\alpha\approx0.5$.  Then
\Eq~\ref{eq:energy} can be simplified by moving to a new density
variable.  Writing
\be
\rho=\rho_0(\rvec,t)w^{-1/(2-\alpha)},
\label{eq:wdef}
\ee
we obtain
\be
\dot{w}=(2-\alpha)\left({\dot{\rho_0}\over\rho_0}-
{1\over\gamma}{\dot{P}\over P}\right)w-
(2-\alpha){\gamma-1\over\gamma}{n_0^2\Lambda(T_0)\over P}.
\label{eq:wdot}
\ee
If the adiabatic compression term is removed by setting
$P\propto\rho_0^\gamma$, then the energy equation takes a particularly
simple form, $\dot{w}=$constant.  However, a more useful
choice is to take $\rho_0\propto\bar{\rho}$.
From the final term of \Eq~\ref{eq:mass}, we see that at high density
when cooling is dominant, then $\dot{\rho}\rho f\sim\,$constant.  This
motivates the substitution
\be
f={(2-\alpha)\over\rho_0}w^{(4-\alpha)/(2-\alpha)}g(w,\rvec,t).
\ee
Then, using \Eqs~\ref{eq:mass}, \ref{eq:energy} and \ref{eq:wdot} we
obtain the following equation for the covariant derivate of $g$ (\ie
following the fluid flow):
\be
{\dot{g}\over g}+(3-\alpha){\dot{\rho_0}\over\rho_0}-
{2-\alpha\over\gamma}{\dot{P}\over P}+\nabla.\uvec=0.
\ee

\end{document}